\documentclass[
    reprint,
    aip,
    amsmath,
    amssymb,
    superscriptaddress]{revtex4-2}
\usepackage{graphicx} 
\usepackage{dcolumn}  
\usepackage{bm}       
\usepackage{amsfonts}
\usepackage{dsfont}
\usepackage{amsmath}
\usepackage{url}

\usepackage{ulem}
\usepackage{color}

\begin{document}

\title{Nonlinear dynamics of electromagnetic field and valley polarization
in WSe$_{2}$ monolayer}
\author{Arqum Hashmi}
\affiliation{Kansai Photon Science Institute, National Institutes for Quantum Science
and Technology (QST), Kyoto 619-0215, Japan}
\author{Shunsuke Yamada}
\affiliation{Center for Computational Sciences, University of Tsukuba, Tsukuba
305-8577, Japan}
\author{Atsushi Yamada}
\affiliation{Center for Computational Sciences, University of Tsukuba, Tsukuba
305-8577, Japan}
\author{Kazuhiro Yabana}
\affiliation{Center for Computational Sciences, University of Tsukuba, Tsukuba
305-8577, Japan}
\author{Tomohito Otobe}
\email[]{otobe.tomohito@qst.go.jp}

\affiliation{Kansai Photon Science Institute, National Institutes for Quantum Science
and Technology (QST), Kyoto 619-0215, Japan}
\date{\today}
\begin{abstract}
Linear and nonlinear optical response of WSe$_{2}$ monolayer is investigated
by two-dimensional Maxwell plus time-dependent density functional
theory with spin-orbit interaction. By applying the chiral resonant
pulses, the electron dynamics along with high harmonic generation
are examined at weak and strong laser fields. WSe$_{2}$ monolayer
shows the linear optical response at the intensity I = 10$^{10}$~W/cm$^{2}$
while a complex nonlinear behavior is observed at I = 10$^{12}$~W/cm$^{2}$.
The nonlinear response of WSe$_{2}$ monolayer in terms of saturable
absorption is observed at strong laser field. By changing the chirality
of the resonant light, a strong circular dichroic effect is observed
in the excited state population. A relatively weak laser field shows
effective valley polarization while strong field induces spin-polarized
carrier peak between $K$($K'$) and $\mathit{\Gamma}$-point via
nonlinear process. On the other hand, the strong laser field shows
high harmonics up to the 11th order. Our results demonstrate that
circularly polarized resonant pulse generate high harmonics in WSe$_{2}$
monolayer of order 3n$\pm1$.
\end{abstract}
\maketitle

\section{INTRODUCTION}

Light-matter interaction is a complex phenomenon that is characterized
by two different spatial scales, the micrometer scale for the wavelength
of the light, and the sub-nanometer scale for the dynamics of electrons\ \cite{hertzog2019strong}.
The interaction of laser light with solids is a promising tool to
investigate the fundamental physical aspects of material properties
in condensed matter physics\ \cite{armstrong1962interactions,fox2002optical,dressel2002electrodynamics}.
In recent days, a tremendous amount of research is focused on the
structural, electronic, and optical properties of thin materials\ \cite{novoselov2005two,tan20202d}.
Thin materials of thickness less than the wavelength of light display
very attractive electronic and optical properties\ \cite{britnell2013strong,kats2016optical}.
In this regard, two-dimensional (2D) layer systems such as graphene\ \cite{novoselov2004electric,novoselov2012roadmap},
transitional metal dichalcogenides (TMDs)\ \cite{wang2012electronics,chhowalla2013chemistry},
and phosphorene\ \cite{xia2014rediscovering} are receiving extensive
research efforts. Among mentioned 2D materials, TMDs are of particular
interest due to broken inversion symmetry and strong spin-orbit coupling
(SOC) for valleytronics and spintronics applications. In addition,
there is a growing interest to study the nonlinear optical properties
of TMDs such as ultrafast carrier dynamics and high order harmonics\ \cite{autere2018nonlinear,langer2018lightwave,liu2017high,saynatjoki2017ultra}.

Time-dependent density-functional theory (TDDFT) has been proposed
as a tool to describe the electron dynamics induced by a time-dependent
external potential from the first principles\ \cite{runge1984density,yabana1996time,ullrich2011time}.
TDDFT has been most successful in the linear response regime to describe
electronic excitation\ \cite{yabana2006real,laurent2013td}. It has
also been applied to the nonlinear and nonperturbative dynamics of
electrons induced by an intense ultrashort laser pulse\ \cite{nobusada2004high,castro2004excited,otobe2008first,shinohara2010coherent}.
The effectiveness of electron dynamics calculations based on TDDFT
is further enhanced by combining it with the Maxwell equations for
electromagnetic fields of pulsed light\ \cite{yabana2012time}. The
combined scheme known as Maxwell-TDDFT formalism has been successfully
applied to crystalline solids\ \cite{lee2014first,sato2015time}.
The Maxwell-TDDFT scheme is quite comprehensive but a high computational
cost limits its applicability. For sufficiently thin materials such
as monatomic layers in which macroscopic electromagnetic fields may
be treated as spatially uniform within the material, we have developed
an efficient approximate method that we call 2D Maxwell-TDDFT scheme
~\cite{yamada2018time,yamada2021determining}. Furthermore, SOC with
non-collinear local spin density is incorporated to extend the usefulness
of the method to study spin-dependent properties.

In this letter, we study the linear and nonlinear optical response
of WSe$_{2}$ monolayer by using a circularly polarized driving field.
WSe$_{2}$ monolayer is chosen to study because of strong SOC and
its most promising phenomena of valleytronics. Laser intensity dependence
of the valley polarization and light propagation in terms of the transmitted
and reflected high harmonics is also investigated by the 2D Maxwell-TDDFT
scheme.

\section{THEORETICAL FORMALISM}

Time-dependent calculations are done using an open-source TDDFT package
Scalable Ab initio Light-Matter simulator for Optics and Nanoscience
(SALMON). Full details of the SALMON code and its implementation are
described elsewhere\ \cite{noda2019salmon,Salmon:Online}. Here,
we briefly describe the 2D Maxwell-TDDFT method for light propagation
in thin layers at normal incidence.

We assume that a thin layer is in the $xy$ plane and a light pulse
propagates along the $z$ axis. By using the Maxwell equations, we
can describe the propagation of the macroscopic electromagnetic fields
in the form of the vector potential ${\bf A}(z,t)$ as, 
\begin{equation}
\left(\frac{1}{c^{2}}\ \frac{{\partial}^{2}}{{\partial t}^{2}}-\ \frac{{\partial}^{2}}{{\partial z}^{2}}\right){\bf A}\left(z,t\right)=\ \frac{4\pi}{c}{\bf J}(z,t),\label{5}
\end{equation}
where ${\bf J}(z,t)$ is the macroscopic current density of the thin
layer.

For a very thin layer, we may assume the macroscopic electric field
inside the thin layer is spatially uniform. We also approximate the
macroscopic electric current density in Eq. (\ref{5}) as 
\begin{equation}
{\bf J}\left(z,t\right)\approx\delta(z){\bf J}_{{\rm 2D}}(t),\label{6}
\end{equation}
where ${\bf J}_{{\rm 2D}}(t)$ is 2D current density (current per
unit area) of the thin layer. We deal with it as a boundary value
problem where reflected (transmitted) fields can be determined by
the connection conditions at $z=0$. From Eq.~(\ref{6}), we obtain
the continuity equation of ${\bf A}(z,t)$ at $z=0$ as follows 
\begin{equation}
{\bf A}(z=0,t)={\bf A}^{{\rm (t)}}(t)={\bf A}^{{\rm (i)}}(t)+{\bf A}^{{\rm (r)}}(t),\label{7}
\end{equation}
where the ${\bf A}^{{\rm (i)}}$, ${\bf A}^{{\rm (r)}}$, and ${\bf A}^{{\rm (t)}}$
are the incident, reflected, and transmitted fields, respectively.
From Eq.~(\ref{5}) and Eq.~(\ref{6}), we get the basic equation
of the 2D approximation method, 
\begin{equation}
\frac{d{\bf A}^{{\rm (t)}}}{dt}=\frac{d{\bf A}^{{\rm (i)}}}{dt}+\ 2\pi{\bf J}_{{\rm 2D}}\left[{\bf A}^{{\rm (t)}}\right](t).\label{8}
\end{equation}
Here, ${\bf J}_{{\rm 2D}}\left[{\bf A}^{{\rm (t)}}\right](t)$ is
the 2D current density that is determined by the vector potential
at $z=0$ and it is equal to ${\bf A}^{{\rm (t)}}(t)$. Time evolution
of electron orbitals in a unit cell of the 2D layer driven by ${\bf A}^{{\rm (t)}}(t)$
is determined by the time-dependent Kohn-Sham (TDKS) equation. By
using the velocity gauge, the TDKS equation for the Bloch orbital
$u_{b,{\bf k}}({\bf r},t)$ (which is a two-component spinor. $b$
is the band index and ${\bf k}$ is the 2D crystal momentum of the
thin layer) is described as,

\begin{equation}
\begin{split}i\hbar\frac{\partial}{\partial t}u_{b,{\bf k}}({\bf r},t)=\Big[\frac{1}{2m}{\left(-i\hbar\nabla+\hbar{\bf k}+\frac{e}{c}{\bf A}^{{\rm (t)}}(t)\right)}^{2}\\
-e\varphi({\bf r},t)+\hat{v}_{{\rm NL}}^{{{\bf k}+\frac{e}{\hbar c}{\bf A}^{{\rm (t)}}(t)}}+{v}_{{\rm xc}}({\bf r},t)\Big]u_{b,{\bf k}}({\bf r},t),
\end{split}
\label{1}
\end{equation}
where the scalar potential $\varphi({\bf r},t)$ includes the Hartree
potential from the electrons and the local part of the ionic pseudopotentials and we have defined  $\hat{v}_{{\rm NL}}^{{\bf k}}\equiv e^{-i{\bf k}\cdot{\bf r}}\hat{v}_{{\rm NL}}e^{i{\bf k}\cdot{\bf r}}$. 
Here, $\hat{v}_{{\rm NL}}$ and ${v}_{{\rm xc}}({\bf r},t)$ are the
nonlocal part of the ionic pseudopotentials and exchange-correlation
potential, respectively. 
The SOC is incorporated through the $j$-dependent nonlocal potential $\hat{v}_{{\rm NL}}$ \cite{theurich2001self}.
The Bloch orbitals $u_{b,{\bf k}}({\bf r},t)$
are defined in a box containing the unit cell of the 2D thin layer
sandwiched by vacuum regions. The 2D current density ${\bf J}_{{\rm 2D}}\left[{\bf A}^{{\rm (t)}}\right](t)$
in Eq.~(\ref{8}) is derived from the Bloch orbitals as follows:
\begin{equation}
\begin{split}{\bf J}_{{\rm 2D}}(t)=-\frac{e}{m}\int dz\int_{\Omega}\frac{dxdy}{\Omega}\sum_{b,{\bf k}}^{{\rm occ}}u_{b,{\bf k}}^{\dagger}({\bf r},t)\\
\times\left[-i\hbar\nabla+\hbar{\bf k}+\frac{e}{c}{\bf A}^{{\rm (t)}}(t)+\frac{m}{i\hbar}\left[{\bf r},\hat{v}_{{\rm NL}}^{{{\bf k}+\frac{e}{\hbar c}{\bf A}^{{\rm (t)}}(t)}}\right]\right]u_{b,{\bf k}}({\bf r},t),
\end{split}
\end{equation}
where $\Omega$ is the area of the 2D unit cell and the sum is taken over the occupied orbitals in the ground state.
In the 2D Maxwell-TDDFT method, coupled Eq.~(\ref{8}) and Eq.~(\ref{1})
are simultaneously solved in real time.

In the weak field limit, we can consider a linear response formalism
for the 2D approximation method. The constitutive relation in linear
response in the form of the 2D current density ${\bf J}_{{\rm 2D}}\left[{\bf A}^{{\rm (t)}}\right](t)$
can be described as follows: 
\begin{equation}
\begin{split}{J}_{{\rm 2D},\alpha}(t)=\sum_{\beta}\int^{t}{dt'}\tilde{\sigma}_{\alpha\beta}\left(t-t'\right)E_{\beta}^{{\rm (t)}}\left(t'\right)\\
=-\frac{1}{c}\sum_{\beta}\int^{t}dt'\tilde{\sigma}_{\alpha\beta}\left(t-t'\right)\frac{dA_{\beta}^{{\rm (t)}}(t')}{dt'},
\end{split}
\label{9}
\end{equation}
where $\alpha$ and $\beta$ are the spatial indices and we have introduced
$\tilde{\sigma}_{\alpha\beta}(t-t')$ as the 2D electric conductivity
of the thin layer. The frequency-dependent 2D conductivity can be
obtained by taking the Fourier transformation of $\tilde{\sigma}_{\alpha\beta}(t)$.
For weak field limit, we can simply get the relation between the incident,
reflection, and transmission by $\tilde{\sigma}_{\alpha\beta}\left(\omega\right)$.
For example, for a given incident field ${\bf A}^{(i)}(t)$, the transmitted
field ${\bf A}^{(t)}(t)$ can be constructed by solving the following
equation in frequency space, instead of solving Eq.~(\ref{8}) in
time.

\begin{equation}
\sum_{\beta}\left(\delta_{\alpha\beta}+\frac{2\pi}{c}\tilde{\sigma}_{\alpha\beta}\left(\omega\right)\right)A_{\beta}^{{\rm (t)}}(\omega)=A_{\alpha}^{{\rm (i)}}(\omega).\label{10}
\end{equation}

WSe$_{2}$ monolayer is used to validate our method, Fig. \ref{fig:1}(a)
illustrates the crystal structure. The lattice constant is set to
$a=b=3.32$~{\AA} We solve the TDKS equation in
the slab geometry with the distance between layers of 20~{\AA}.
The dynamics of the 24 (12 electrons for W, and 12 electrons for Se)
valence electrons are treated explicitly while the effects of the
core electrons are considered through norm-conserving pseudopotentials
from the OpenMX library\ \cite{morrison1993nonlocal}. The adiabatic local density approximation
with Perdew-Zunger functional\ \cite{perdew1992accurate} is used
for the exchange-correlation. We adopt a
spin noncollinear treatment for the exchange-correlation potential\cite{von1972local,oda1998fully}.
  The spatial grid sizes and k-points
are optimized according to the convergent results. The determined
parameter of the grid size is 0.21~{\AA} while the
optimized k-mesh is 15 \texttimes{} 15 in the 2D Brillouin zone. The
time step size is set to 5\texttimes 10$^{-4}$ femtosecond (fs).
We consider the circularly polarized laser field with a time profile
of ${\cos}^{4}$ envelope shape for the vector potential, which gives
the electric field of the applied laser pulse through the following
equation, 
\begin{multline}
{\bf A}^{{\rm (i)}}\left(t\right)=-\frac{cE_{{\rm max}}}{\omega}\cos^{4}\left(\pi\frac{t-T_{P}/2}{T_{P}}\right)\\
\left[\hat{{\bf x}}{\cos}\left\{ \omega\left(t-\frac{T_{P}}{2}\right)\right\} +\hat{{\bf y}}{\sin}\left\{ \omega\left(t-\frac{T_{P}}{2}\right)\right\} \right],\\
(0<t<T_{p})\label{12}
\end{multline}
where$\ \omega$ is the average frequency, $E_{{\rm max}}$ is the
maximum amplitude of the electric field and $T_{P}$ is the pulse
duration. We use the frequency of 1.55~eV, and $T_{p}$ is set to
30~fs and the computation is terminated at 50~fs.

\section{RESULTS AND DISCUSSION}

\begin{figure}[h]
\centering \includegraphics[scale=0.5]{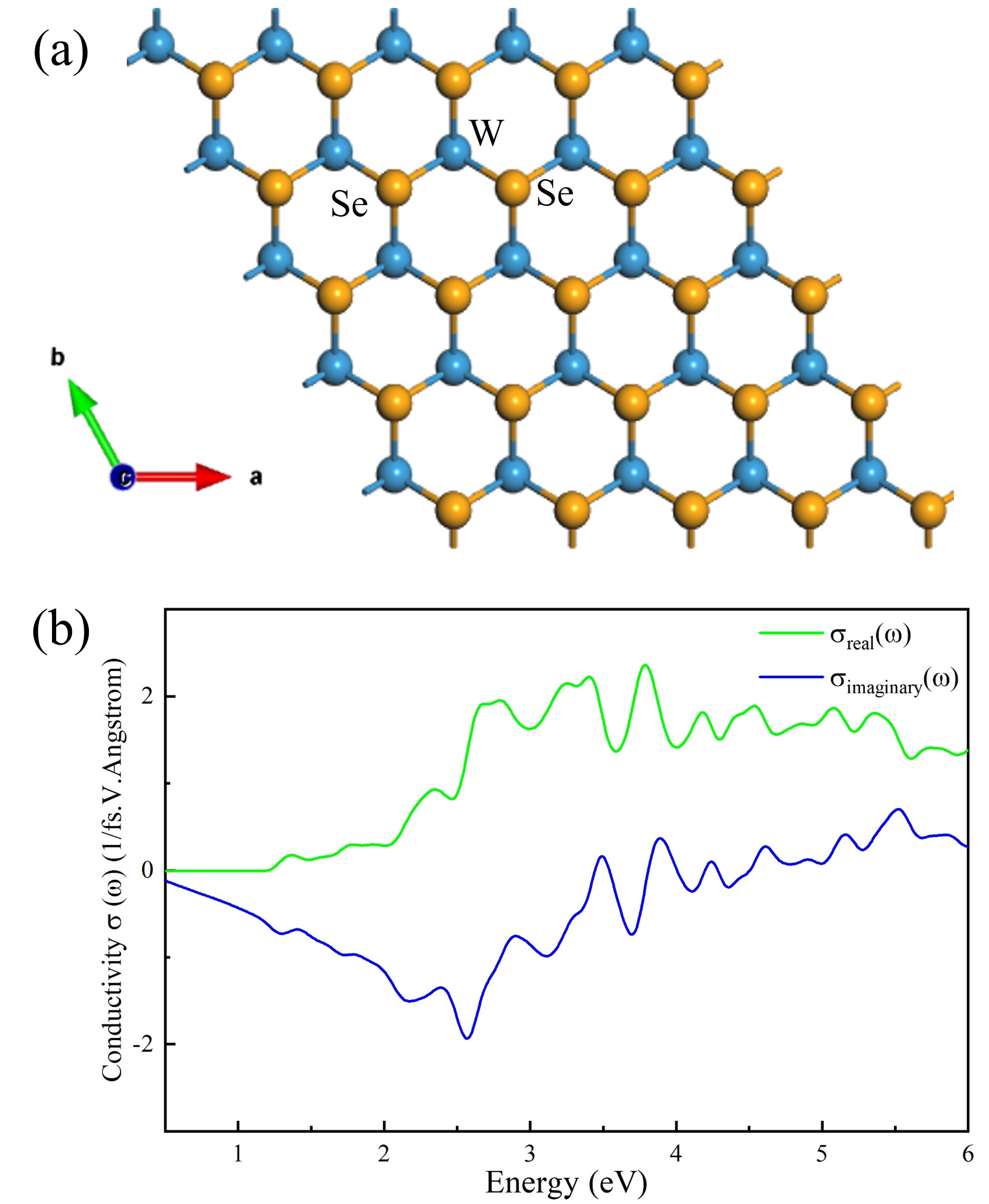} \caption{(a) Crystal structure of WSe$_{2}$ in top view. (b) Frequency-dependent
optical conductivity of WSe$_{2}$ monolayer.}
\label{fig:1} 
\end{figure}

2D WSe$_{2}$ monolayer has strong SOC and all the calculations are
performed by taking the SOC effect into consideration. The calculated
bandgap is \ensuremath{\sim} 1.25~eV. The Zeeman-type spin splitting
due to SOC is also checked. The value of spin splitting for valence
band maxima (VBM) is \ensuremath{\sim} 450~meV, consistent with
previous works\ \cite{kormanyos2015k,rasmussen2015computational,le2015spin}.
Fig. \ref{fig:1}(b) depicts the real and imaginary parts of the optical
conductivity $\tilde{\sigma}(\omega)$ as a function of the photon
energy. $\tilde{\sigma}(\omega)$ provides an understanding
of the band structure properties, the linear dependence of $\tilde{\sigma}(\omega)$
at small energies with zero value at \ensuremath{\omega}=0 reveal
the pure semiconducting nature of WSe$_{2}$ monolayer. The peaks
at higher energies in the optical conductivity correspond to the interband
transition from the valence band to the conduction band.

\begin{figure*}
\centering\includegraphics[scale=0.45]{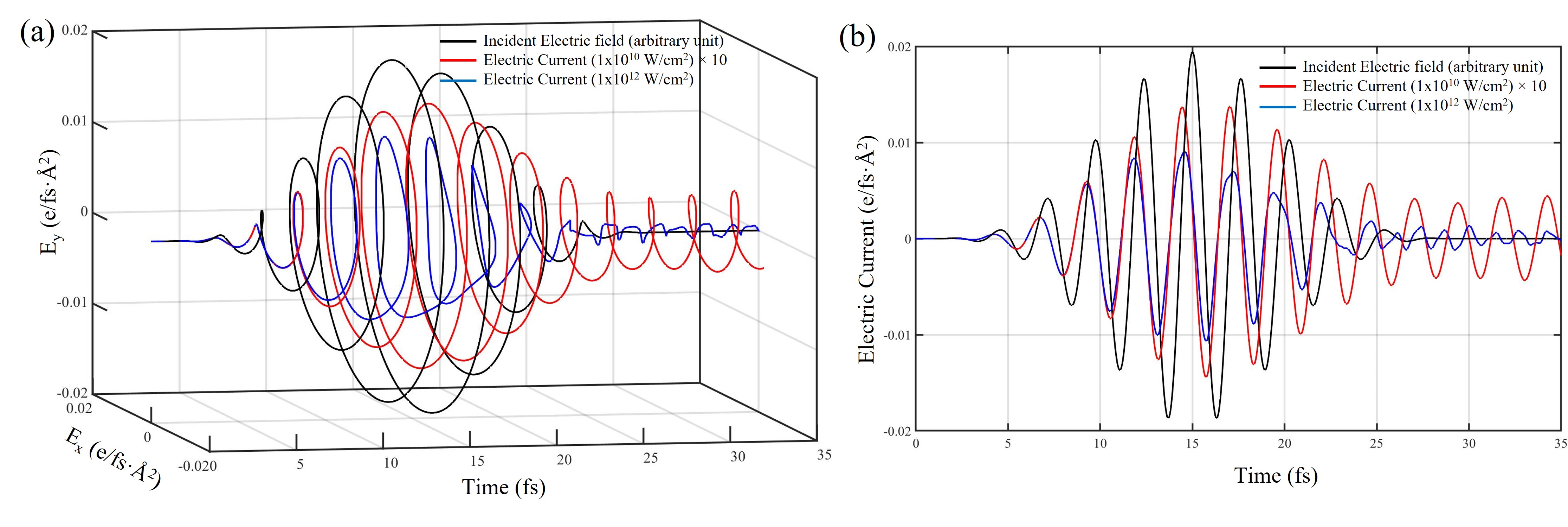} \caption{Typical time evolution calculation for WSe$_{2}$ monolayer. (a) Applied
electric field waveform and induced electric current density at weak
and strong laser field. (b) X-Y view of current density in (a).}
\label{fig:2} 
\end{figure*}

Fig. \ref{fig:2} shows a typical electron dynamics calculation of
the WSe$_{2}$ monolayer. The time profile of the incident circular
polarized electric field and the induced electric current at weak
and strong intensity is shown in Fig. \ref{fig:2}(a). The current
at weak intensity is multiplied by a factor of 10 so that the difference
between two currents indicates the nonlinear effect at the strong
intensity. For a more clear comparison, we have also shown the X-Y
view of the current density in Fig. \ref{fig:2}(b). At the weak intensity,
the temporal evolution of the induced current mostly follows the driving
laser profile, indicating that a linear optical response of WSe$_{2}$
monolayer dominates at I = 10$^{10}$~W/cm$^{2}$. We also note that
there appears substantial current after the incident pulse ends, since
the frequency of the applied laser pulse is above the bandgap value.
At the strong intensity, the current is initially very close to the
case of the weak intensity. However, the current gradually becomes
weaker than that expected from the linear response. We consider that
this nonlinear effect suppressing the current comes from the saturable
absorption that often appears in various 2D materials ~\cite{wang2013ultrafast,liu2021application,uemoto2021first}.
The saturable absorption takes place by two mechanisms: a decrease
of electrons in the valence band by the excitation, and an increase
of electrons in the conduction band that block the excitation from
the valence band. We also note that the current after the incident
pulse ends is much weaker than the case of weak intensity. This indicates
that nonlinear effect works to cancel the coherence that produced
the delayed current in the linear case.

\begin{figure}[h]
\centering\includegraphics[scale=0.32]{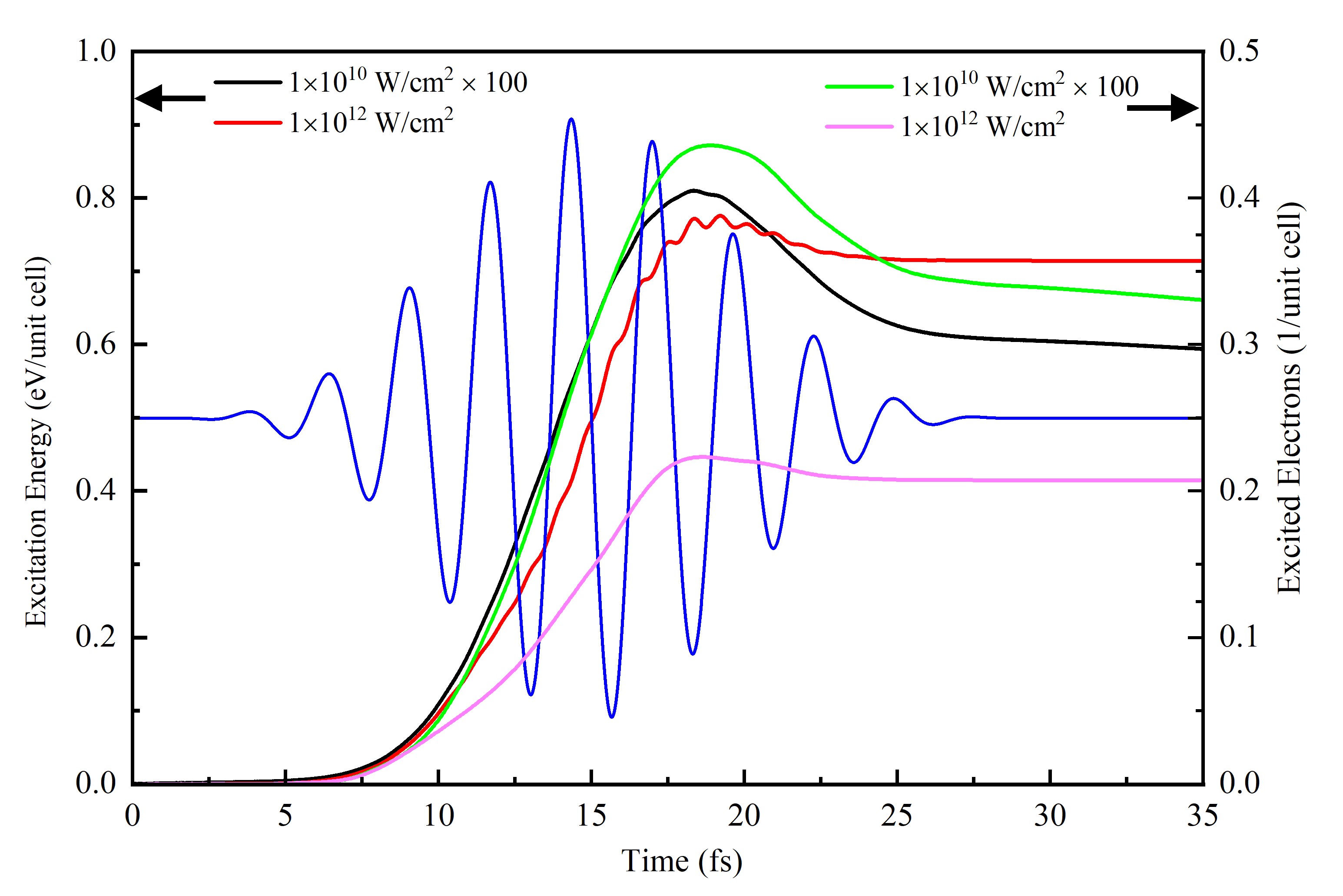} \caption{Temporal development of excitation energies and excited electrons
against incident vector potential. For comparison, the results for
weak intensity (I = 10$^{10}$~W/cm$^{2}$) is scaled up by a factor
of 100.}
\label{fig:3} 
\end{figure}

We show the calculated results of electric excitation energy and the
number density of excited electrons in Fig. \ref{fig:3}. To see the
nonlinear effects, we have scaled up the excitation by a factor of
100 for weak intensity. Note that the weak and strong field curves
should coincide and show similar behavior if the response is linear.
We first consider the case of weak intensity. The electronic excitation
energy and the number of excited electrons show a similar time profile
with a peak around 18~fs and then decrease. The ratio of the two
curves is about 1.8~eV and is close to the photon energy of 1.55~eV,
confirming the dominance of the one-photon absorption process. The
decrease continues even after the incident pulse ends at 30~fs. This
decrease is due to the emission of light that is described in the
present 2D Maxwell-TDDFT scheme, and is related to the appearance
of the current after the pulse ends in Fig. \ref{fig:2}. We next
move to the case of strong intensity. The number of excited electrons
is much smaller than the estimation by linear excitation mechanism.
This is due to the saturable absorption, as we discussed for the current
shown in Fig. 2. However, the electronic excitation energy looks to
show mostly linear behavior. We consider that this is due to a cancellation
of two nonlinear effects, saturable absorption and multi-photon absorption.
Although the number of excited electrons is suppressed by the saturable
absorption, the excited electrons can absorb multiple photons, as
is evaluated from the ratio of electronic excitation energy and the
number of excited electrons which is about 3.8~eV, more than twice
of the energy of the incident photon.

\begin{figure*}
\centering \includegraphics[scale=0.45]{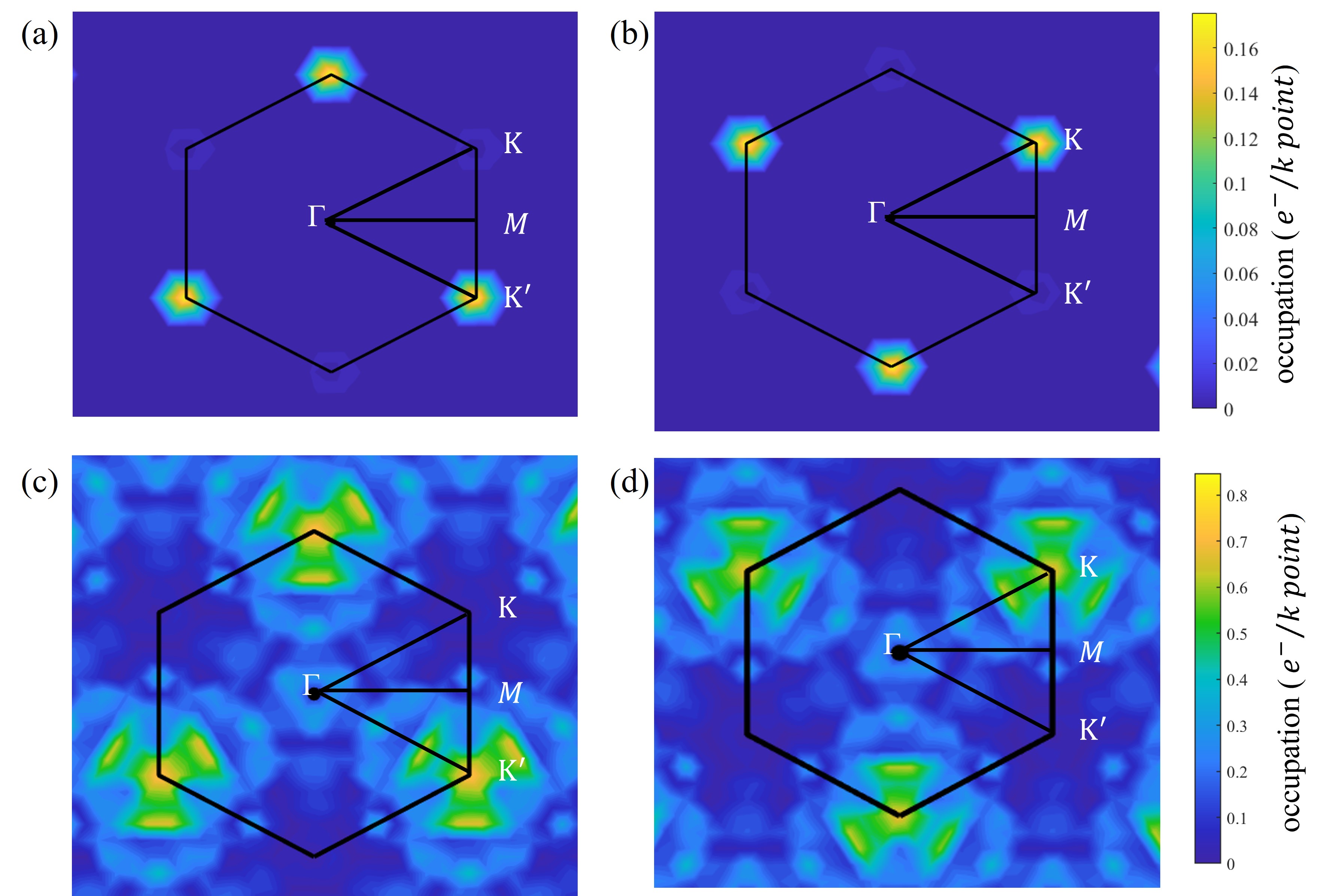} \caption{Distribution of k-resolved electron populations in the first BZ of
the conduction band at the end of the pulse (a) Right hand and (b)
Left hand circularly polarized light at I = 10$^{10}$ W/cm$^{2}$.
(c, d) Same as the upper panel but at I = 10$^{12}$ W/cm. Electron
population is summed over the entire conduction band }
\label{fig:4} 
\end{figure*}

\begin{figure*}
\centering \includegraphics[scale=0.45]{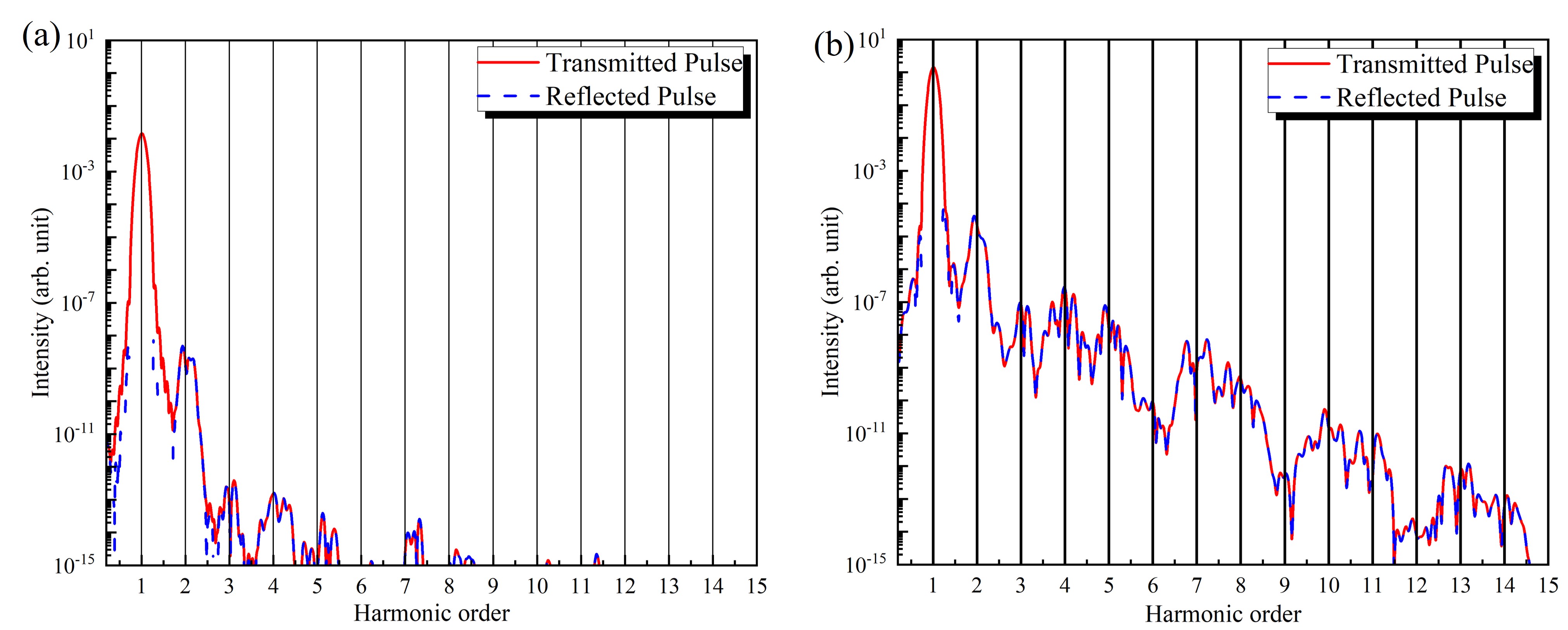} \caption{HHG spectra of WSe$_{2}$ monolayer. High harmonic signals of the
reflected and the transmitted wave at (a) I = 10$^{10}$~W/cm$^{2}$
(b) I = 10$^{12}$~W/cm$^{2}$.}
\label{fig:5} 
\end{figure*}

To get an understanding of valley pseudospin, we investigate the k-resolved
excited electron populations. It is well known that the responses
of the conductance for spin-dependent electrons to left- and right-handed
resonant circularly polarized light are just the opposite for TMDs\ \cite{zaumseil2014electronic,zhang2014electrically,xie2016manipulating}.
Fig. \ref{fig:4} shows the excited carrier population by right (\ensuremath{\sigma}$^{+}$)
and left (\ensuremath{\sigma}$^{-}$) circular helicities. Valley
polarization results are shown in Fig. \ref{fig:4}(a, b) for I =
10$^{10}$~W/cm$^{2}$ and for at I = 10$^{12}$~W/cm$^{2}$ in
Fig. \ref{fig:4}(c, d). A strong circular dichroic effect is evident
in the excited state population both at weak and strong intensity.
The valley degeneracy is lifted and pumping with \ensuremath{\sigma}$^{+}$
polarized light excites electrons in the $K$ valley while pumping
with \ensuremath{\sigma}$^{-}$ polarized light excites the $K^{'}$
valley, so-called valley-spin locking. Due to nonlinear interaction
at I = 10$^{12}$~W/cm$^{2}$, the excited electrons are not localized
around $K$/$K^{'}$ and start to spread in the Brillouin zone and
a new peak between K (K') and $\mathit{\Gamma}$-point appears with
nonlinear process. The new peak induced by the strong field may explain
the behavior of electric current density at I = 10$^{12}$~W/cm$^{2}$
as shown in Fig. \ref{fig:2}. The excited electron at different points
of the Brillouin zone indicates the electronic current with different
phase and frequency. This phase difference decreases the total current
after the pulse ends. Overall, the k-resolved excited electron populations
demonstrate that the valley polarization state is quite robust against
the field strength.

High harmonic generation (HHG) is a promising tool to study strong-field
effects and ultrafast electron dynamics in 2D materials. Experimentally,
one can primarily measure the fraction of reflected or transmitted
pulses of the light, and the 2D Maxwell-TDDFT scheme used for our
calculations can describe the propagation of the light wave dynamics
as well. Fig. \ref{fig:5} shows the HHG spectra included the reflected
(${\bf A}^{(r)}(t)$) and transmitted (${\bf A}^{(t)}(t)$) pulses.
Fig. \ref{fig:5}(a) show the HHG spectra for a peak intensity of
I = 10$^{10}$ W/cm$^{2}$ while Fig. \ref{fig:5}(b) is for I = 10$^{12}$~W/cm$^{2}$.
First of all, HHG spectra of the reflected and transmitted pulses
completely coincide with each other except for the fundamental frequency,
as is understood from the continuity equation of Eq.~(\ref{7}).
We find that a resonant circularly polarized driver (\ensuremath{\omega}
= 1.55~eV) generates strong high harmonics. For the weak intensity
of 10$^{10}$~W/cm$^{2}$, the spectrum includes very weak HHG components
except for the second harmonic. It was as expected because the electronic
current density shows the linear optical response at this intensity.
On the other hand, the strong intensity of 10$^{12}$~W/cm$^{2}$ (Fig.
\ref{fig:5}(b)) shows the harmonic order up to 11th order. WSe$_{2}$
monolayer displays both odd and even harmonic peaks. Our results show
that the in-plane harmonics of WSe$_{2}$ monolayer under resonant
circular driver has an order of 3n \textpm{} 1 (where n is an integer),
while the multiple of 3 harmonic orders are suppressed. The three-fold
rotational symmetry of the WSe$_{2}$ monolayer lead to the selection
rule of 3n \textpm{} 1. The 3rd harmonic appears in Fig. \ref{fig:5}(b)
is the exception to this rule and it may be related to the relative
phase effect as explained in Ref.\ \cite{alon1998selection}.

\section{CONCLUSION}

In this work, we have studied the linear and nonlinear optical response
of WSe$_{2}$ monolayer by the classical Maxwell equations combined
with TDKS equations to describe the propagation of electromagnetic
fields in thin 2D layers. By applying the chiral resonant pulses,
the electron dynamics and HHG are shown at a weak and strong laser
field. WSe$_{2}$ monolayer shows the linear optical response at weak
intensity. By changing the chirality of the resonant light, a strong
circular dichroic effect is observed in the excited state population.
A relatively weak laser field shows effective valley polarization
while strong field induces new peak of spin-polarized carrier by nonlinear
process. WSe$_{2}$ monolayer displays the nonlinear optical characteristic
of saturable absorption at strong laser field. The 2D Maxwell-TDDFT
scheme offers a clear advantage to measure the HHG in terms of the
transmitted and reflected waves. HHG spectra of the reflected and
transmitted pulses completely coincide with each other. HHG in WSe$_{2}$
monolayer is observed up to the 11th order and circularly polarized
resonant driver have high harmonics of 3n$\pm1$ order in WSe$_{2}$
monolayer.

\section{AUTHOR INFORMATION}

\noindent \textbf{Corresponding Author}

\noindent {*}E-mail: otobe.tomohito@qst.go.jp

\noindent \textbf{Author Contributions}


\noindent \textbf{Notes}

\noindent The authors declare no competing financial interest.

\section{ACKNOWLEDGMENT}

\noindent This research is supported by JST-CREST under Grant No.
JP-MJCR16N5. This research is also partially supported by JSPS KAKENHI
Grant No. 20H02649, and MEXT Quantum Leap Flagship Program (MEXT Q-LEAP)
under Grant No. JPMXS0118068681 and JPMX0118067246. The numerical
calculations are carried out using the computer facilities of the
Fugaku through the HPCI System Research Project (Project ID: hp210137),
SGI8600 at Japan Atomic Energy Agency (JAEA), and Multidisciplinary
Cooperative Research Program in CCS, University of Tsukuba.

 \bibliographystyle{naturemag}
\bibliography{bibliography}
\eject 
\end{document}